\documentstyle[12pt,colordvi,epsf]{article}
\def\met{\mbox{${\hbox{$E$\kern-0.6em\lower-.1ex\hbox{/}}}_T$}} 

\begin{document}
\begin{titlepage}
\vspace{-0.25in}
\begin{flushleft}
\leftline\small {$11^{th}$ Rencontres de Physique de la Vall\'ee d'Aoste,
March 2--8, 1997}
\end{flushleft}
\vspace{-0.3in}
\begin{flushright}
\large {D\O\ Conf--97--9}
\end{flushright}
\vspace{-0.35truein}
\begin{flushright}
\large {FERMILAB--Conf--97/178--E}
\end{flushright}
\vspace{0.2in}

\begin{center}
{\large\bf RECENT RESULTS FROM THE SEARCH FOR NEW PHENOMENA AT D\O\ } \\
\vspace{2.5cm}
{\large Jay A. Wightman} \\
\vspace{.5cm}
{\sl Iowa State University} \\
{\sl for the D\O\ Collaboration} \\
\vspace{2.5cm}
\vfil
\begin{abstract}

We present results from several new searches for physics beyond the 
Standard Model.  We describe a search based on the scalar sum of the
transverse energy of the event, a global quantity nearly independent of the 
event topology.  We summarize our searches
for first generation leptoquarks into all three decay channels, 
$e q \overline{e} \overline{q}$, $e q \overline{\nu} \overline{q}$, and
$\nu q \overline{\nu} \overline{q}$ and note that this is the first time that 
the triumvirate of decay channels has been searched.  
We do not find any evidence for production of first generation leptoquarks 
and set a lower limit on the mass of the leptoquark of 175 GeV/c$^2$, assuming 
the decay is exclusively into $e q \overline{e} \overline{q}$.  We also present
results from the first search for a third generation leptoquark 
with charge = $\pm 1/3$.
Again, we find no evidence for its existence for a mass less than 80 GeV/c$^2$.
Finally, we discuss one of our searches for supersymmetry, specifically the
pair--production of $\widetilde{e}$, $\widetilde{\nu}$, and 
$\widetilde{\chi_2^0}$ where the decay yields final states with two photons
plus missing transverse energy (\met).  
We set limits on the production cross section 
ranging from 1 pb to 400 fb, depending on the mass.  This analysis also sets a
model--independent limit of 
$\sigma \cdot B( p \overline{p} \rightarrow \gamma \gamma$ + \met + $X) <
185$ fb at the 95\% CL for $E_T^{(\gamma)} > 12$ GeV and $\vert \eta \vert 
< 1.1$ and \met $>25$ GeV.

\end{abstract}

\end{center}
\end{titlepage}

\section{Introduction}

The successes of the Standard Model are legendary and numerous; however, 
the model is not complete as it leaves several questions unanswered, for
example, 
what is the origin of the mass hierarchy and why are there three generations
of fermions, thus opening the door for extensions to the model.
Many extensions to the Standard Model include new, heavy particles 
while others introduce new interactions.
We report new results from a few of the searches presently underway at D\O.  
The first is an analysis of events with large scalar transverse energy where we
are looking for evidence of contact interactions.  
The next topic includes updates to two searches
for first generation leptoquarks decaying to electrons plus quarks and first 
results from searches for first generation leptoquarks decaying to 
neutrinos plus quarks and third generation leptoquarks decaying to neutrinos
plus $b$ quarks.  Finally, we summarize some newly published results from an 
analysis of events with two photons plus missing transverse energy searching
for pair--production of $\widetilde{e}$, $\widetilde{\nu}$, and 
$\widetilde{\chi_2^0}$.  
We note that this is only a small subset of the total package of active 
searches that we are pursuing.

\section{Large Scalar Transverse Energy as a Window on New Physics}

As noted above, many extensions to the Standard Model involve additional, heavy
particles or new interactions.  
The clearest evidence for new particles would be an invariant
mass peak due to on--shell production, but no such peaks have been
found.  Therefore, if new massive particles exist, the mass must be larger
than $\sqrt{\hat s}$, the parton--parton CM energy available at the Tevatron.
Indirect evidence of their existence can then be inferred from an increase in
the cross section with increasing $\sqrt{\hat s}$ over that predicted by the
Standard Model.  
For the case of new interactions the characteristic energy scale is
usually larger than the electro--weak scale.

We are developing a generic search strategy to look for evidence of new 
physics.  We define the quantity 
$$H_T \equiv \sum^N_{i=1} \vert E_T^{(i)} \vert,$$ where $N$
is the number of jets with transverse energy $E_T \geq 20$ GeV (with no
requirement on the jet multiplicity).  Thus, $H_T$ has 
only a weak dependence on 
the event topology.  To demonstrate a ``proof of principle'' of the sensitivity
of the $H_T$ analysis to new physics, we apply it to a specific extension to
the Standard Model, composite quarks.
Previous searches for quark sub--structure used the inclusive jet
cross section.  
Recently, an alternative analysis based on the angular correlation between the
leading two jets (two highest $E_T$ jets) in the event has published a limit
on the 
compositeness scale, $\Lambda^{^*}$, of 1.8 TeV assuming all six flavors of
quarks are composite and destructive interference in the 
Lagrangian\cite{cdf-comp1-ref}.  D\O\ presented a preliminary limit at this
conference\cite{d0-inc-jet-ref}.  Because the $H_T$ analysis uses a more
global quantity it complements the 
inclusive jet and di--jet angular correlation measurements.

$H_T$ has the following advantages.  It is the best measure we have of the 
transverse component of $\sqrt{\hat s}$ since it sums over most of the jets
in the event, omitting only the low--$E_T$ jets where the reconstruction 
efficiency begins to drop, the jet energy scale is not well--determined, and
underlying event uncertainties are large.  
One of the problems inherent in any jet analysis is the details of the jet
algorithm used to define the jets.  For example, when using a cone algorithm
which employs a fixed cone size, one question that arises is how to resolve two
nearby energy clusters.  If they cannot be resolved then they are merged to 
produce a single jet, and if they can be resolved they are split with some
prescription for how to partition the energy.  This decision of merge/split can
easily populate/depopulate the high energy regime of the inclusive jet cross
section, exactly where one expects to find evidence of new physics.  For the
cross section as a function of $H_T$, this is not a concern so long as the jet
energy scale is well--behaved.  Final--state radiation can also depopulate the 
high energy regime of the inclusive jet cross section when a high--$E_T$ jet
radiates a moderate--$E_T$ jet; once again, we find the $H_T$ distribution 
is robust.

The $H_T$ analysis uses a shape comparison between the measured and
the predicted cross section and is therefore insensitive to the overall 
normalization.  There are several input parameters that are needed for the QCD
calculation, such as choice of parton distribution function and renormalization
scale, which can result both in normalization and shape differences from one 
choice to the next.  We find that the shape of the $H_T$ cross section does not
depend on the choice of renormalization scale.  We show an example of this in
the lefthand plot in Fig. 1 where each curve is the ratio between the 
cross sections for two different
choices of renormalization scale.  The Monte Carlo cross section is generated 
using the NLO generator, Jetrad\cite{jetrad-ref}, with renormalization scale, 
$\mu = 0.5, 1.0, {\rm ~or~} 1.5 \times E_T$ 
where $E_T$ is the transverse energy of the leading jet of the event.  
Here we use the CTEQ2ML\cite{cteq-ref} parton 
distribution functions.  
The small variation in shape of each curve as a function of $H_T$ is due to the
ansatz function (an exponential whose argument is a polynomial in $H_T$) 
used to smooth the Monte Carlo distributions, 
$${{d \sigma}\over{d H_T}} = 
exp \Bigl [ \sum^3_{i=0} (a_i \times H_T^i) \Bigr ],$$ where the $a_i$ are 
coefficients determined in the fit.

\setlength{\unitlength}{0.7mm}
\begin{figure}[h]
\begin{picture}(200,100)(-10,1)
\mbox{\epsfxsize7.0cm\epsffile{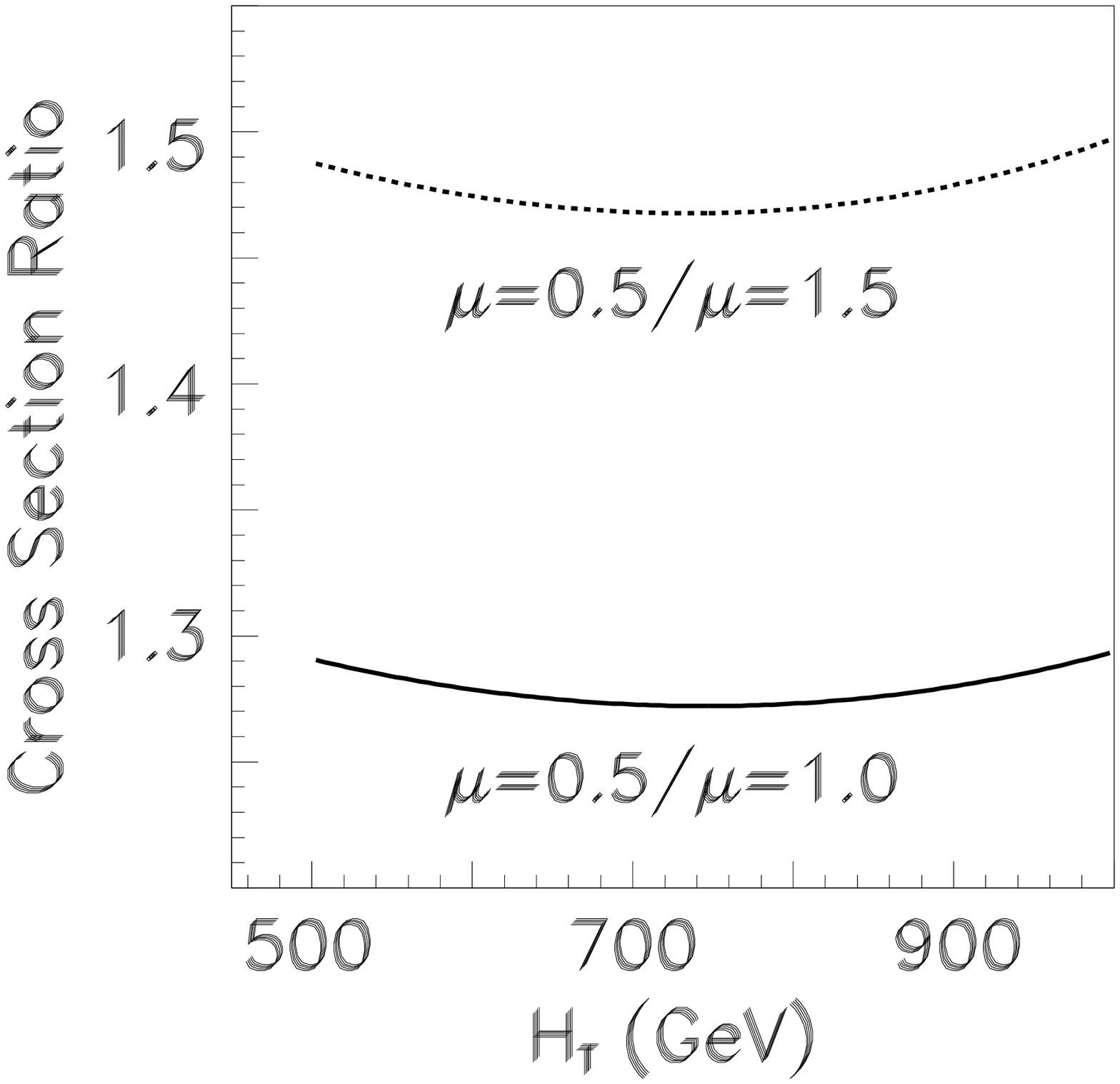}
      \epsfxsize7.0cm\epsffile{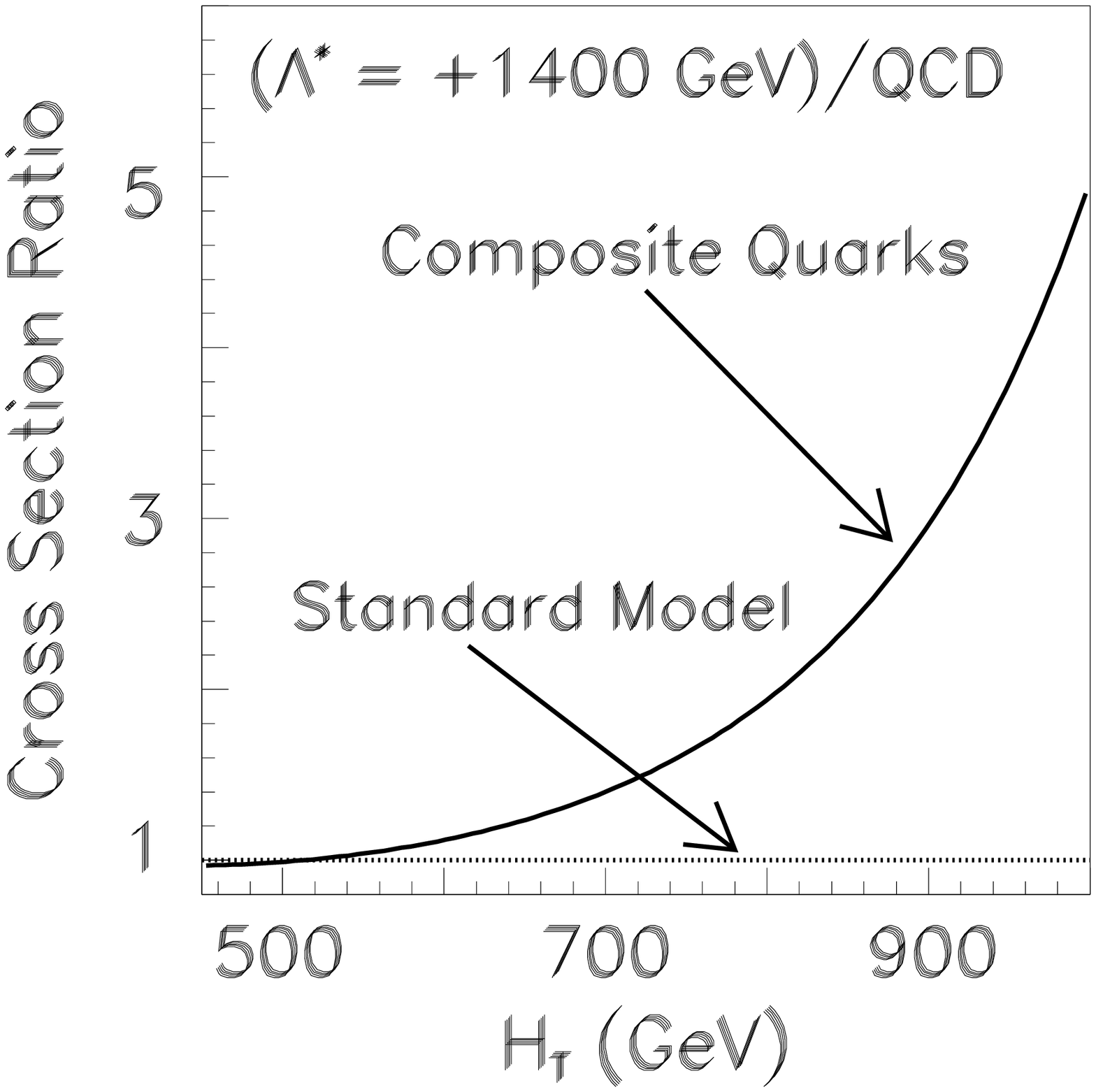}}
\end{picture}
\caption{Lefthand plot: The ratio of NLO QCD cross sections for various 
choices of renormalization scale, $\mu$.  Righthand plot: The ratio of LO
cross sections for composite quarks ($\Lambda^{^*} = 1400$ GeV and destructive
interference in the lagrangian) and QCD.}
\end{figure}

In general, changing the order of the QCD
calculation, from leading--order (LO) to next--to--leading--order (NLO), 
results in both
a normalization as well as a shape change in the cross section.  The more 
inclusive quantity experiences less shape change, an important consideration
since the models for new processes are implemented as LO processes in the
event generators.  Another important aspect of doing a shape analysis is that
it minimizes some of the systematic errors, such as the uncertainty in the jet
energy scale, and eliminates others, such as the uncertainty in the integrated
luminosity.

The model for quark sub--structure that we employ is 
from Eichten, {\it et al.},\cite{rmp-ref} with all six quarks allowed to be 
composite and both signs of the interference term possible.  
This model is implemented in the LO generator, PYTHIA,
\cite{pythia-ref} but we use for comparison the QCD cross section generated 
with Jetrad.
Therefore, we make the a priori assumption that the ratio between QCD and 
quark compositeness generated at LO is identical to what
we would find at NLO if the model was implemented there.  This ratio we term
the K--factor; an example is shown in the righthand plot in Fig. 1.  
We apply this ratio to the QCD
cross section generated by Jetrad to simulate quark compositeness at NLO.  For
our Monte Carlo event generation we use 
$\mu = 0.5 \times E_T$ of the leading jet and the CTEQ3M parton distribution
functions, consistent with the D\O\ inclusive jet 
analysis\cite{d0-inc-jet-ref}.

Here we report preliminary results from an analysis of 
$90.4 \pm 4.9$ pb$^{-1}$ of
data taken in Run 1b, the 1994--95 run.
The trigger used was a multi--jet trigger which was fully efficient for 
$H_T \geq 500$ GeV.  Because the cross section 
decreases by several orders of magnitude for $500 \leq H_T \leq 1000$ GeV, we 
linearize the comparison with the Monte Carlo generated cross section 
by taking the difference between the
two and normalizing to the Monte Carlo cross section.  
As noted above, we are doing
a shape analysis, so we normalize the generated cross section to 
match the data
in the bin $H_T = 500$ GeV.  The results are shown in Fig. 2 where for the
lefthand plot we generated 
QCD ($\Lambda ^ {^*} = \infty$) and for the righthand plot we generated 
composite quarks with $\Lambda ^ {^*} = 1400$ GeV and destructive interference
in the Lagrangian ($+$ sign of the interference term).  The error bars are
statistical only with the systematic errors due to the jet energy scale shown
as the dotted and dashed lines.  The extraction of a limit on the scale of 
quark sub--structure is still underway, but Fig. 2 indicates that the
data are in good agreement with NLO QCD up to the 
highest energies probed and that the $H_T$ analysis is sensitive to contact 
interactions.  With the new jet energy scale and its
concomittantly smaller uncertainty reported at this 
conference\cite{d0-inc-jet-ref}, we expect to extract a very competitive limit
on the scale of quark compositeness.  Finally, we note that even though we used
quark compositeness as a specific example for the preceding discussion we have
a ``proof of principle'' 
of the sensitivity of the $H_T$ analysis to new interactions at scales much 
higher than the electro--weak scale.

\setlength{\unitlength}{0.7mm}
\begin{figure}[h]
\begin{picture}(200,100)(-10,1)
\mbox{\epsfxsize7.0cm\epsffile{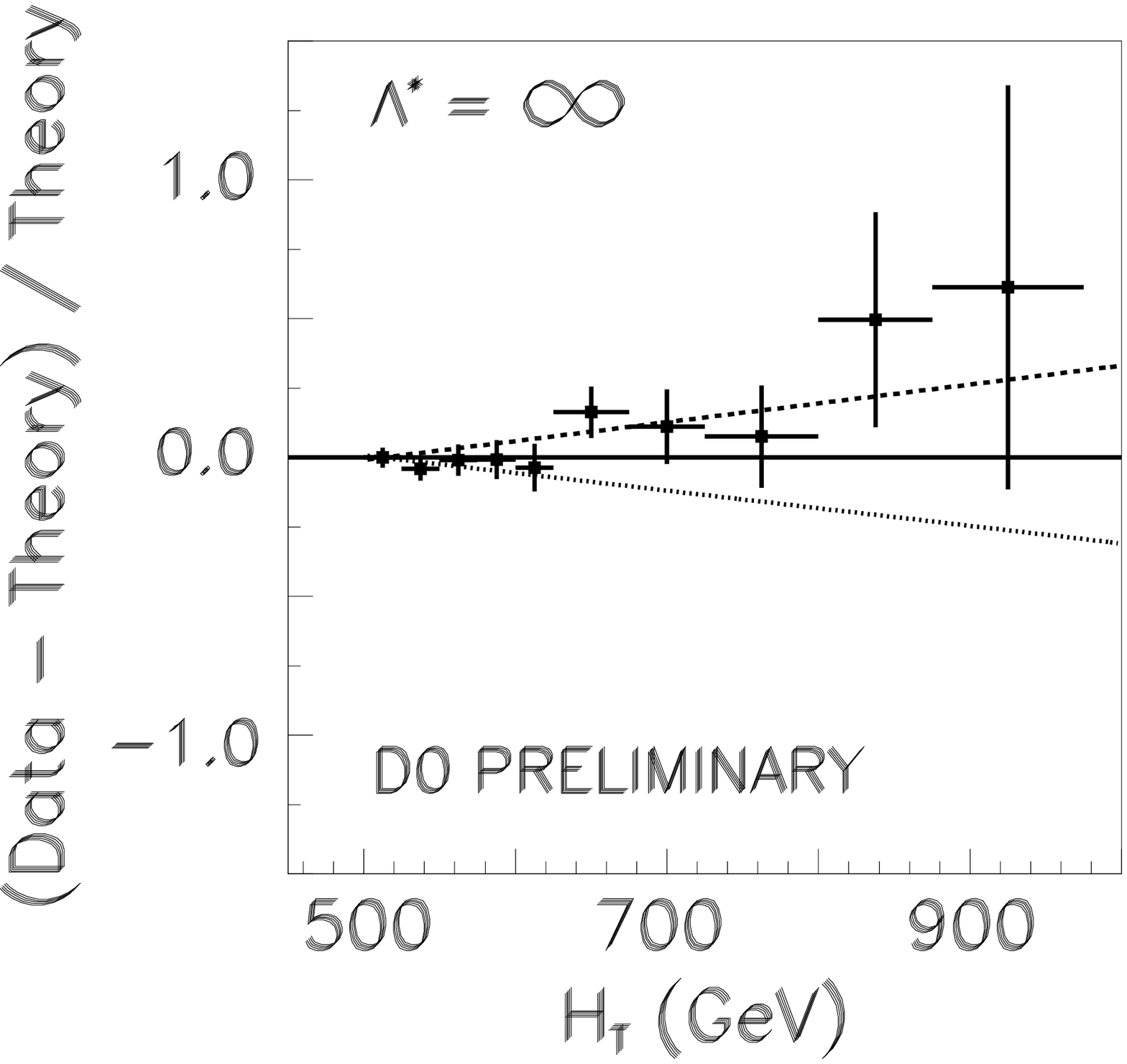}
      \epsfxsize7.0cm\epsffile{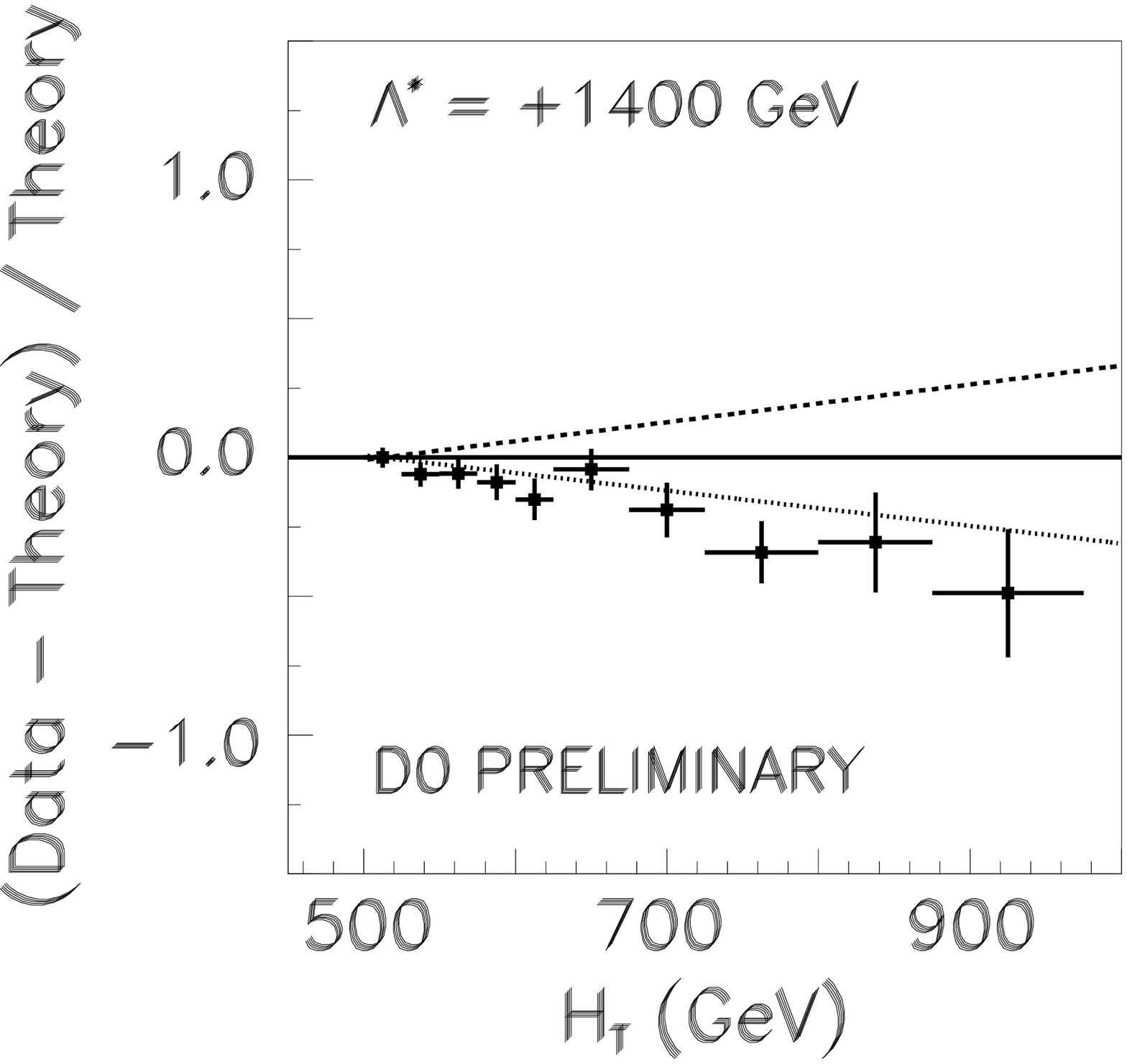}}
\end{picture}
\caption{Lefthand plot: Data comparison with NLO QCD, where theory refers to
the Monte Carlo cross section generated using Jetrad, see text for details.  
Righthand plot: Data comparison with NLO quark 
sub--structure ($\Lambda^{^*} = 1400$ GeV and destructive interference in
the Lagrangian).  The error bars are statistical, and the error band is the 
systematic uncertainty due to the jet energy scale.}
\end{figure}

\section{Leptoquarks}

Leptoquarks are particles that carry both lepton number and color, and, 
therefore, they couple both to leptons and quarks.  These arise in many 
extensions to the Standard Model\cite{lq-ref}.  They couple with
an unknown coupling strength which is usually parameterized in terms of the 
electro--weak coupling $$g^2 = 4 \pi \alpha k,$$ where $\alpha$ is the fine
structure constant and $k$ is an unknown constant.  We find that we are fully
efficient for $k \geq 10^{-12}$.  
This cutoff arises from the requirement that the leptoquark decays within
the D\O\ beam pipe.  
Rare decay experiments set strict limits on contributions from flavor--changing
neutral currents which translate into very high limits on the masses of the
leptoquarks.  
If we require that the leptoquarks couple to a 
single Standard Model generation only, 
these limits are considerably lower allowing direct searches
at present day colliders.  There is 
renewed interest in leptoquark searches at the Tevatron following the recent
report of an excess of events at high--$Q^2$ by the H1\cite{h1-ref} and 
ZEUS\cite{zeus-ref} collaborations at DESY.  
Leptoquarks are produced at the Tevatron via strong pair--production.  
They can be either scalar or vector
particles; we report here on searches for scalar leptoquarks.  
We define $\beta$ as the branching fraction of the decay of the 
leptoquark to the charged lepton, $\ell$, plus a quark.  Three basic final 
states arise: (1) both leptoquarks decay to $\ell q$ with branching fraction,
$\beta^2$; (2) one leptoquark decays to $\ell q$ and the other to $\nu q$, with
branching fraction, $2 \beta (1 - \beta)$; (3) both leptoquarks decay to 
$\nu q$, with branching fraction $(1 - \beta)^2$.

\subsection{First Generation Leptoquarks -- $e q \overline{e} \overline{q}$ 
channel}

For this analysis, we select events that have two high--$E_T$ electrons ($E_T >
25$ GeV) and two 
high--$E_T$ jets ($E_T > 30$ GeV), consistent with the event topology.  
The electrons must pass stringent quality cuts\cite{d0-top-ref}.  To 
reduce the background from $Z$ + jets events, we veto events where the 
di--electron mass lies within $\pm 15$ GeV/c$^2$ of the nominal mass of the
$Z$.  The overall efficiency for detecting first generation leptoquarks in
this channel depends on the mass of the leptoquark and ranges from 0.21\% for
a mass of 40 GeV/c$^2$ to 24.1\% for a mass of 250 GeV/c$^2$.  The analysis
is optimized for a leptoquark mass of 160 GeV/c$^2$.  
The full Run 1 data sample ($117.7 \pm 6.4$ pb$^{-1}$) yields 3 events
which pass our selection criteria.

Five sources of events contribute to the background for this
analysis.  The largest is Drell--Yan production of $Z/\gamma$ + jets.
The second largest contribution is from  
$t\overline{t}$ pair--production with subsequent decay to 
di--electron final states.  Smaller contributors to the background are
$Z$ + jets, where 
$Z \rightarrow \tau \overline{\tau} \rightarrow e\overline{e}$, 
$W^+W^-$ + jets, where $W^+W^- \rightarrow \overline{e} \nu e \overline{\nu}$, 
and QCD multi--jet events where two of the jets
fluctuate to mimic electrons.  We predict $2.9 \pm 1.1$ events from these
backgrounds.

Our search is a null search since the number of events that pass our analysis
criteria in the data are well described by the predicted background.  
Therefore, we interpret the results as a 95\% CL upper limit on the cross 
section as a function of the mass of the leptoquark.  An example of such a 
limit is presented in Fig. 3 where we have assumed $\beta = 1.0$ and find 
a lower limit on the mass of 175 GeV/c$^2$.  The
theoretical cross section is a LO calculation from Ref. \cite{lepto-xsec-ref};
this calculation yields a smaller cross section than we had used 
previously\cite{d0-lepto-dpf-ref} resulting in a somewhat lower mass limit.

\setlength{\unitlength}{0.7mm}
\begin{figure}[h]
\begin{picture}(100,100)(-60,1)
\mbox{\epsfxsize7.0cm
      \epsffile{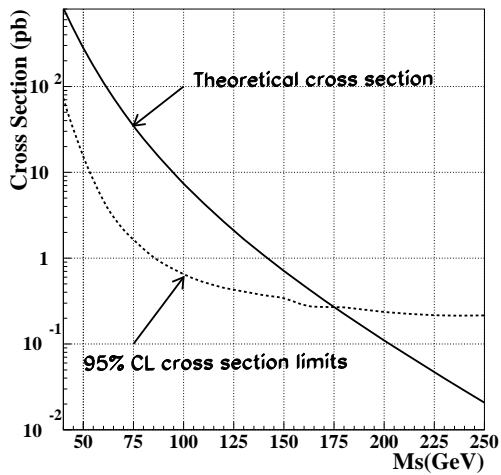}}
\end{picture}
\caption{95\% CL upper limit on the cross section for production of first 
generation scalar leptoquarks as a function of the mass, $\beta = 1.0$.}
\end{figure}

\subsection{First Generation Leptoquarks -- $e q \overline{\nu} 
\overline{q}$ channel}

The event selection requirements for this channel are a single high--$E_T$ 
electron ($E_T > 25$ GeV), two high--$E_T$ jets ($E_T > 25$ GeV), and large
missing transverse energy (\met $> 40$ GeV), consistent with the event 
topology (the \met ~is due to the presence of the neutrino).  To reduce 
the background from $W$ + jets we require the transverse
mass, $m_T$, of the electron and \met ~satisfy 
$m_T > 100$ GeV/c$^2$.  We discriminate against 
background from $t\overline{t}$ pair--production by requiring
$H_T > 170$ GeV, where $H_T \equiv
\sum \vert E_T^{(j)} ( > 15 {\rm ~GeV}) \vert + \vert E_T^{(e)} \vert$, 
and vetoing events containing a reconstructed muon with $p_T > 4$ GeV/c.  
The overall efficiency again varies 
as a function of the leptoquark mass and ranges from 0.03\% for a leptoquark
mass of 40 GeV/c$^2$ to 14.5\% for a leptoquark mass of 250 GeV/c$^2$.  As
with the $eq \overline{e} \overline{q}$ channel, this analysis is optimized for
a leptoquark mass of 160 GeV/c$^2$.
Analysis of $103.7 \pm 5.6$ pb$^{-1}$ of data yields 3 events passing the 
requirements outlined above.

There are two main sources of background.  
The first is 
$t\overline{t}$ pair--production with subsequent decay to a single electron in
the final state; this is the largest residual source of background.  
The second is $W$ plus two jets, where $W \rightarrow
e \overline{\nu}$.  
Both of these sources of Standard Model background exhibit
large missing transverse energy.  
We predict $4.0 \pm 1.1$ events from these sources.

Once again, the number of data events passing our selection criteria is 
well--modeled by the background.  We interpret our null search result as a
95\% CL upper limit on the cross section as a function of the leptoquark mass. 
An example of this is shown in Fig. 4 where we have assumed $\beta = 0.5$ and 
find a lower limit on the mass of 132 GeV/c$^2$.  The
theoretical cross section is from the calculation of Ref. 
\cite{lepto-xsec-ref}.

\setlength{\unitlength}{0.7mm}
\begin{figure}[h]
\begin{picture}(100,100)(-60,1)
\mbox{\epsfxsize7.0cm
      \epsffile{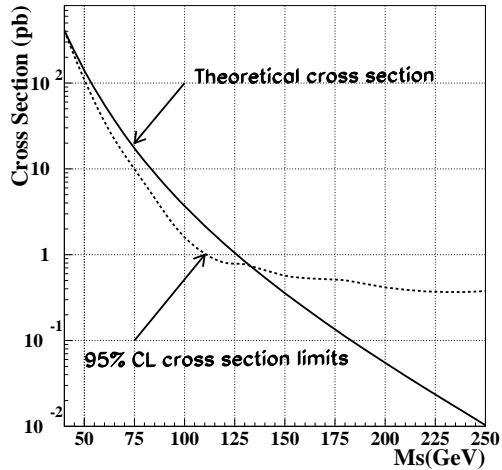}}
\end{picture}
\caption{95\% CL upper limit on the cross section for production of first 
generation scalar leptoquarks as a function of the mass, $\beta = 0.5$.}
\end{figure}

\subsection{First Generation Leptoquarks -- $\nu q \overline{\nu} 
\overline{q}$ channel}

The preliminary analysis of this channel completes the 
triumvirate of searches for first generation leptoquarks and is the first 
result from this channel.  While this is in
actuality a mixed--generational search since we are not able to tag the flavor
of the neutrinos, we note that this 
analysis is valid for all three generations as long as we obtain a null result.
We also stress that the analysis we present here is not optimized to 
search for first generation leptoquarks, rather it is a simple application 
of our previously published light top squark analysis\cite{d0-stop-ref} 
to this search.

The event requirements are two high--$E_T$ jets ($E_T > 30$ GeV) and large 
missing transverse energy (\met $> 40$ GeV).  To reduce the vector boson 
contribution to the background, we veto events containing charged leptons.  
To minimize ambiguity in the \met ~determination we also require only single
interactions which yields an effective integrated luminosity of $7.4 \pm 0.4$
pb$^{-1}$ from Run 1a, the 1992--93 run.  
We find 3 events surviving our selection criteria.  The main source
of background is vector bosons + jets for which we predict $3.5 \pm 1.2$ 
events.  Once again, we find the number of events passing our event selection
is well--modeled by the background resulting in a null search.  We interpret 
this null result as a 95\% CL upper limit on the cross section as a function
of the leptoquark mass.  We present this limit in Fig. 5 where $\beta = 0.0$
and find a lower limit on the mass of 71 GeV/c$^2$.
The theoretical cross section is calculated from Ref. \cite{lepto-xsec-ref}.

\setlength{\unitlength}{0.7mm}
\begin{figure}[h]
\begin{picture}(100,100)(-50,10)
\mbox{\epsfxsize7.0cm\epsffile{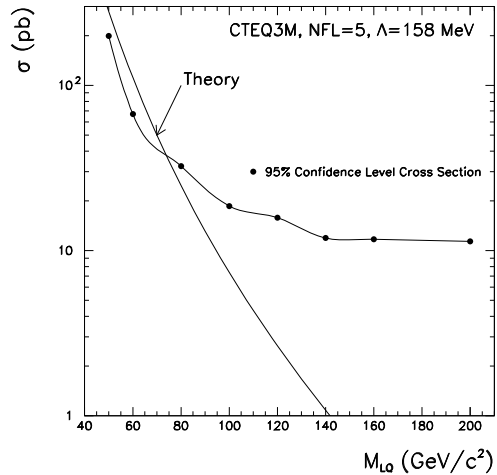}}
\end{picture}
\caption{95\% CL upper limit on the cross section for production of first  
generation scalar leptoquarks as a function of the mass, $\beta = 0.0$.}
\end{figure}

\subsection{First Generation Leptoquarks -- combined channels}

We then vary $\beta$ over the allowable range $0.0 \leq \beta \leq 1.0$ for the
three analyses presented above to produce limits of leptoquark mass
versus $\beta$ and 
combine all three channels to determine the limit on first generation 
leptoquark production as a function of $\beta$ as shown in Fig. 6.  We
also include in Fig. 6 for comparison the LEP I direct search limit of 
45 GeV/c$^2$ and our previously published limit\cite{d0-lepto-ref}.  For 
$\beta = 1.0$ we find a lower limit on the first generation scalar 
leptoquark mass of 175 GeV/c$^2$, while $\beta = 0.5$ and $0.0$ yield lower
limits of 147 and 71 GeV/c$^2$, respectively.  

\setlength{\unitlength}{0.7mm}
\begin{figure}[h]
\begin{picture}(100,100)(-60,1)
\mbox{\epsfxsize7.0cm
     \epsffile{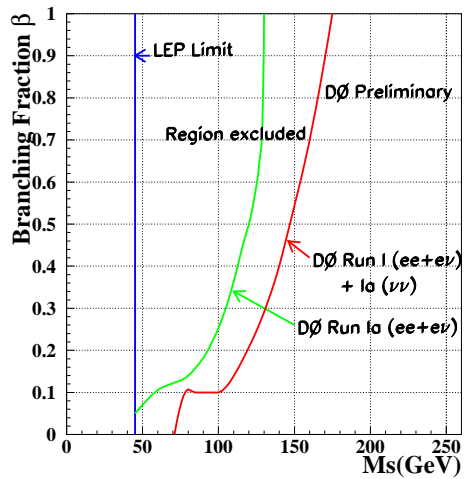}}
\end{picture}
\caption{First generation leptoquark mass limit versus $\beta$.
The present limit contour is labelled D\O\ Run 1 ($ee + e\nu$) + 1a 
($\nu \nu$).}
\end{figure}

\subsection{Third Generation Leptoquarks -- $\nu b \overline{\nu} 
\overline{b}$ channel}

The search for third generation leptoquarks decaying to 
$\nu b \overline{\nu} \overline{b}$ is optimized for 
a leptoquark with mass less than that of the top quark.
As opposed to the searches for first generation leptoquarks where both 
quark species were kinematically allowed in the final state, here we are
restricted to $b$ quarks only.  Therefore, this search is not a search in terms
of $\beta$ but rather is a search for a leptoquark with charge = $\pm 1/3.$

The event selection requires large missing transverse energy, \met $> 35$ GeV,
and two jets, where one or both of the jets is tagged as a $b$ quark jet due
to the presence of a $\mu$--tag, consistent with the event topology.  
For a tagged jet, we require $E_T > 10$ GeV,
while untagged jets must satisfy $E_T > 25$ GeV.  We apply additional 
topological constraints to enhance the $b$ quark decays 
of the leptoquarks and reduce 
mismeasurement sources of \met.  To enhance the $b$ quark 
decays we require that the di--jet system carry most of the energy 
since $\mu$'s from $b$ quark decays tend to be softer 
than $\mu$'s from $W$ decay.  
We veto events where there is too much energy in an annular cone around the
tagging muon thus reducing 
contributions from gluon splitting, $g \rightarrow b \overline{b}$.  
Finally, to reduce 
mismeasurement sources of \met, we require that the \met ~vector not be aligned
or anti--aligned with either of the two leading jets.  The effective total 
acceptance varies as a function of the leptoquark mass and ranges from 
0.41\% for a mass of 60 GeV/c$^2$ to 3.4\% for a mass of 150 GeV/c$^2$.  
Analysis of $85.6 \pm 4.6$ pb$^{-1}$ of data yields 2 events that satisfy the
event requirements outlined above.

The sources of background for this channel are Standard Model sources of 
$b$ quarks, such as $t \overline{t}$ pair--production, $W/Z \rightarrow b$, 
and QCD $b$ quark events.  The QCD events are difficult to estimate 
with an expectation of $4 \pm 4$, so we
arbitrarily set this contribution to zero resulting in the most conservative 
limit we can set.  We predict
$2.8 \pm 0.7$ background events from the other two sources.  The number of
events passing our selection criteria is consistent with the background, so
we interpret the null result as a 95\% CL upper limit on the production cross
section as a function of the leptoquark mass.  The limit is plotted in Fig. 7
where we find for third generation leptoquarks with charge = $\pm 1/3$ 
a lower limit on the mass 
of 80 GeV/c$^2$.  We include in Fig. 7 the LEP I direct search limit of
45 GeV/c$^2$.  The theoretical cross section shown in Fig. 7 is
based on our earlier calculation\cite{d0-lepto-dpf-ref}.  

\setlength{\unitlength}{0.7mm}
\begin{figure}[h]
\begin{picture}(100,100)(-60,1)
\mbox{\epsfxsize7.0cm\epsffile{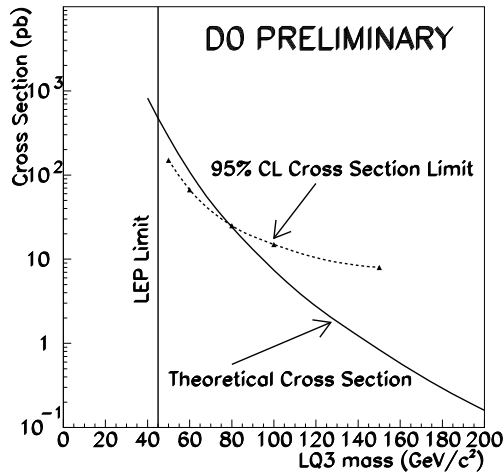}}
\end{picture}
\caption{95\% CL upper limit on the cross section for production of third  
generation scalar leptoquarks with charge = $\pm 1/3$ as a 
function of the mass.}
\end{figure}

\section{Supersymmetry}

Supersymmetry is one of the most elegant extensions to the Standard Model since
it generalizes the space--time symmetry without altering the gauge symmetry.
It also provides the framework for unification with gravity which occurs at a 
sufficiently high enough mass scale, like the Planck scale.  The Minimal
Supersymmetric Standard Model, MSSM, consists of adding a super--partner to 
every Standard Model particle and results in a plethora of undetermined 
parameters, such as the masses and mixings of the particles.  

Recent models of low--energy supersymmetry suggest signatures involving one or
more photons plus missing transverse energy\cite{susy-models-ref}.  One 
feature common to these models is the prediction of tens of events in the 
$>100$ pb$^{-1}$ of data taken in Run 1.  The photon plus \met ~final state
arises from pair--production of $\widetilde{e}$, $\widetilde{\nu}$, and 
$\widetilde{\chi_2^0}$, where the $\widetilde{e}$ and $\widetilde{\nu}$ decay
to $\widetilde{\chi_2^0}$ which subsequently decays to $\widetilde{\chi_1^0}$ 
and a photon.  This analysis is recently 
published\cite{d0-eegg-ref}, and we summarize the search for these 
signatures.  The event requirements are two central ($\vert \eta \vert < 1.1$) 
photons with $E_T > 12$ GeV and large missing transverse energy (\met $> 25$
GeV).  To minimize the contribution from $Z \rightarrow ee$ where the two 
electrons are reconstructed as photons, we veto events with a di--photon 
invariant mass within $\pm 10$ GeV/c$^2$ of the nominal $Z$ mass.  To reduce
the background from radiative $W$ decays ($W(e \nu)\gamma$ and 
$W(e \nu \gamma)$) we require an azimuthal separation of the photons ($\Delta
\phi_{\gamma \gamma} > 90^{\circ}$).  Analysis of $93.3 \pm 11.2$ pb$^{-1}$ 
of data yields no event satisfying the selection criteria.

The largest source of background comes from QCD multi--jet events where two of
the jets are misidentified as photons and mismeasurement provides the large
\met.  Lesser sources of background are processes like 
$W \rightarrow e\overline{\nu}$, $\tau \rightarrow eX$, and 
$t\overline{t} \rightarrow eX$, where 
the $e$ is misidentified as a photon, in combination with either a photon 
or a jet misidentified as a photon.  We predict $2.0 \pm 0.9$ 
background events in our event sample.  In the lefthand plot of Fig. 8 
we show the \met ~distribution measured in the event; the 
points are the background prediction, and the data are shown in the histogram.
As can be seen from this plot, there is no visible excess in the 
data.  Detecting zero events with \met $> 25$ GeV 
is consistent with the background model, so we interpret
this null result as a 95\% CL upper limit on the production cross section which
is shown in the righthand plot of Fig. 8 as a function of the 
neutralino mass difference.  We find that the mean photon $E_T$ and the mean
\met ~of these events is typically given by the neutralino mass difference.
We examine several masses for the pair--production of
$\widetilde{e}$, $\widetilde{\nu}$, and $\widetilde{\chi_2^0}$.  The cross
section limit ranges from 1 pb to 400 fb for neutralino mass differences 
$>20$ GeV/c$^2$, almost independent of the species and mass being 
pair--produced.

\setlength{\unitlength}{0.7mm}
\begin{figure}[h]
\begin{picture}(200,100)(-10,10)
\mbox{\epsfxsize7.0cm\epsffile{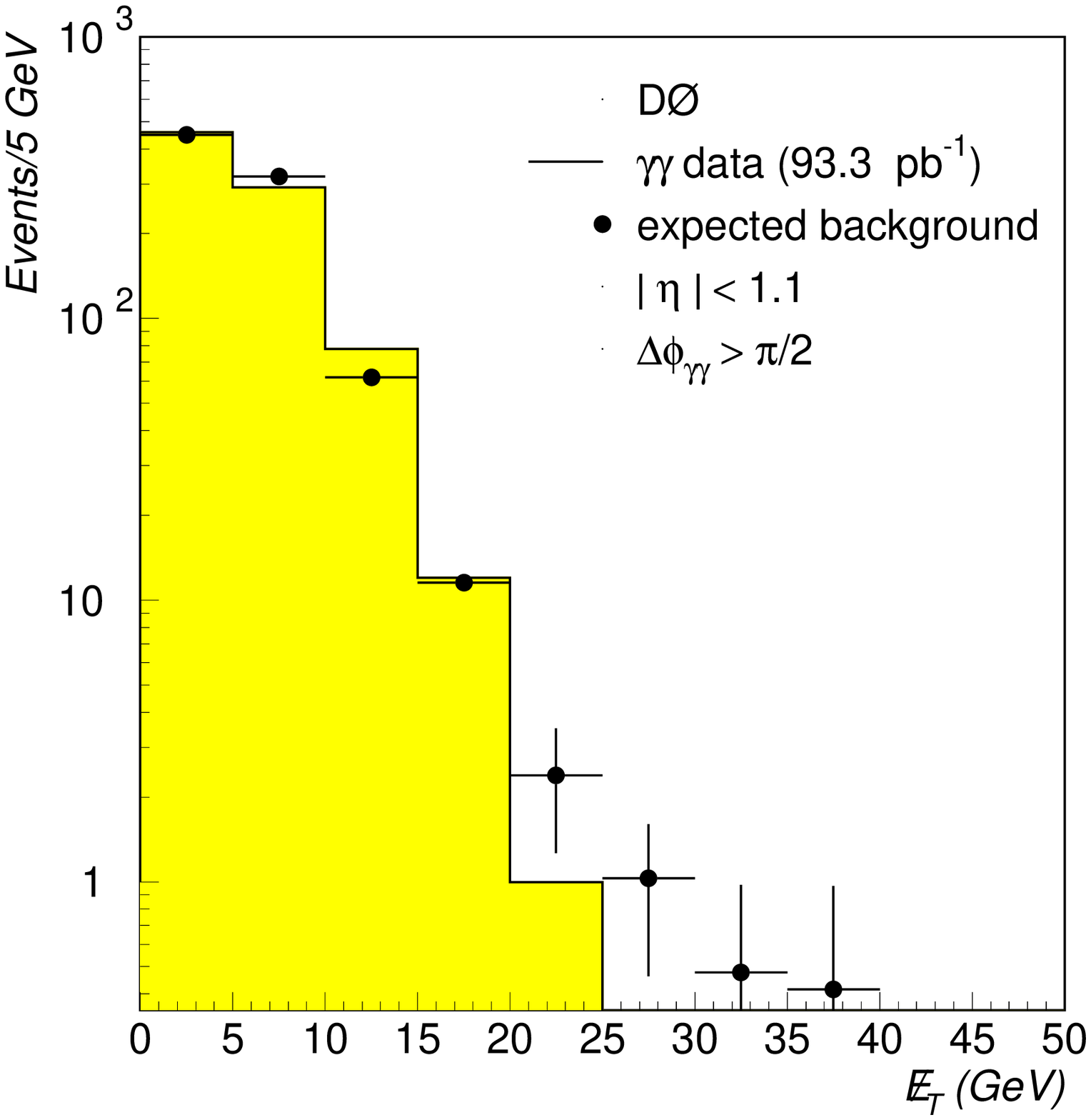}
      \epsfxsize7.0cm\epsffile{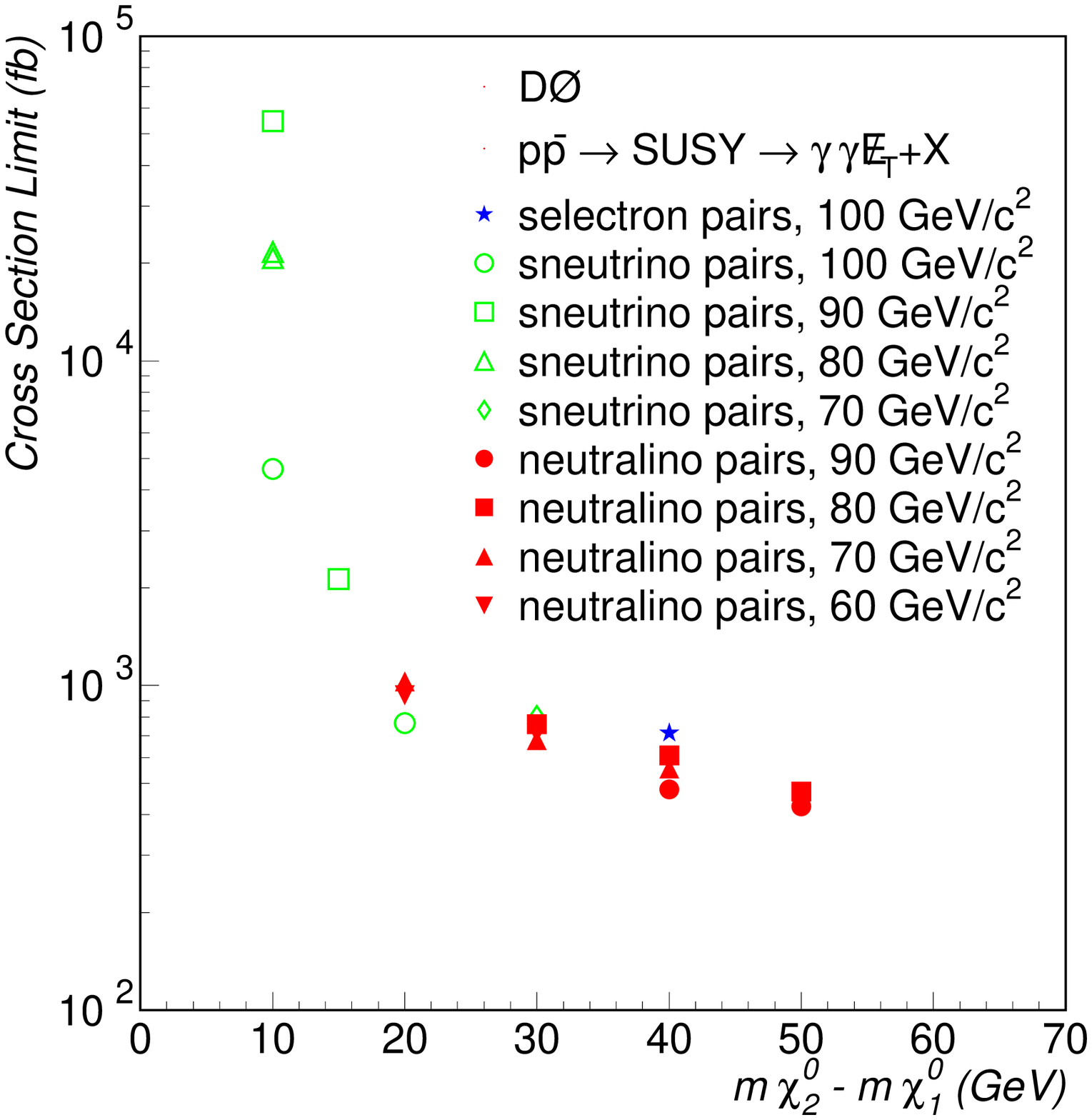}}
\end{picture}
\caption{Lefthand plot: The \met ~distribution for $\gamma \gamma$ data 
(histogram) and background (black circles).  Righthand plot: 95\% CL upper
limits on the pair--production cross section where the decay 
$\widetilde{\chi_2^0} \rightarrow \gamma + \widetilde{\chi_1^0}$ has been
forced.}
\end{figure}

The weak dependence of the cross section limit on the neutralino mass 
difference above 20 GeV/c$^2$ arises from the 
near independence of the event efficiency (trigger plus reconstruction) for 
the kinematic requirements above.  Therefore, we extend the analysis to set 
a model--independent upper limit on the cross section for final 
states satisfying our event 
selection criteria of two central ($\vert \eta \vert < 1.1$)
photons with $E_T > 12$ GeV and \met $>25$ GeV of
$\sigma \cdot B( p \overline{p} \rightarrow \gamma \gamma$ + \met + $X) <
185$ fb at the 95\% CL.  The reason that this model--independent limit is 
better than the supersymmetry search limit is that typically 
only 25--50\% of the
supersymmetric events satisfy our kinematic requirements.  The effect of this
model--independent limit on some of the recently proposed models is to exclude
a considerable fraction of their parameter space.

\section{Summary and Future Prospects}

We have shown preliminary results from the $H_T$ analysis and have argued
the efficacy of this analysis for searching for physics beyond the Standard
Model.  We use quark compositeness as our example of new physics to 
describe the search.  We have presented preliminary results and 
find that the cross section as a function of $H_T$ is
consistent with NLO QCD up to the highest energies probed.   
We do not set a limit on the compositeness scale at this time as we are still
working on this phase of the analysis; however, we do find that this analysis
is sensitive to contact interactions, as required in the quark sub--structure
model.  The future for this analysis is very bright as we incorporate the new
jet energy scale and begin to map out the rich array of proposed extensions to
the Standard Model.

We have presented new results from our search for 
first generation leptoquarks which, for the first time,
covers the triumvirate of decay channels 
($eq \overline{e} \overline{q}$, $eq \overline{\nu}
\overline{q}$, and $\nu q \overline{\nu} \overline{q}$).  
The results from the two decay channels
containing an electron are updates to what we have previously shown.  The 
analysis employs a new calculation of the production cross section which has
resulted in a somewhat decreased mass limit.  We have presented the first
results from an analysis of the $\nu q \overline{\nu} \overline{q}$ 
channel.  Even with a non--optimized analysis of a limited data set, we 
set a limit of 71 GeV/c$^2$ on the mass of the leptoquark for $\beta = 0.0$.
Combining all three channels we find lower limits on the mass of 
first generation leptoquarks of 175 GeV/c$^2$ ($\beta = 1.0$),
147 GeV/c$^2$ ($\beta = 0.5$), and 71 GeV/c$^2$ ($\beta = 0.0$).
This is not the last word we 
will have to say regarding the search for first generation leptoquarks as
we plan to re--optimize the analysis for higher leptoquark masses, such as 200 
GeV/c$^2$.  We are also exploring invariant mass constraints on the $eq$ pairs
in both the $eq \overline{e} \overline{q}$ and $eq \overline{\nu} \overline{q}$
channels.  We are examining a NLO cross section calculation which will result
in higher mass limits.  We will be adding the entire Run 1 data set to the 
$\nu q \overline{\nu} \overline{q}$ analysis and optimizing for searching for
leptoquarks.  Finally, we are searching
for vector leptoquarks, where the production cross section is 
substantially larger than for scalar leptoquarks.

We have presented the first results from a search for a 
third generation scalar leptoquark with charge = $\pm 1/3$ 
decaying to $\nu b \overline{\nu} 
\overline{b}$.  We find a lower limit on the leptoquark mass of 80 GeV/c$^2$.
The future for this
analysis involves re--optimizing for masses greater than the limit quoted 
above, employing the NLO cross section calculation, and searching for vector
leptoquarks.

We have presented results from one of our searches for supersymmetry, the 
pair--production of $\widetilde{e}$, $\widetilde{\nu}$, and 
$\widetilde{\chi_2^0}$.  We force the decays of $\widetilde{e}$ and
$\widetilde{\nu}$ to $\widetilde{\chi_2^0}$ plus a photon with the subsequent
decay of $\widetilde{\chi_2^0}$ to $\widetilde{\chi_1^0}$ plus a photon.  We 
find 95\% CL upper limits on the production cross section that range from 1 pb
to 400 fb for neutralino mass differences $>$ 20 GeV/c$^2$, nearly 
independent of the mass or the species of the particles being pair--produced.
This weak dependence allows us to set a model--independent 95\% CL upper limit
on the cross section for producing any final state satisfying our kinematic
requirements (two central photons with $\vert \eta \vert < 1.1$ and $E_T > 12$ 
GeV plus \met $> 25$ GeV) of 
$\sigma \cdot B( p \overline{p} \rightarrow \gamma \gamma$ + \met + $X) <
185$ fb at the 95\% CL.  The effect of this
model--independent limit on some of the recently proposed models is to exclude
a considerable fraction of their parameter space.

We have reported on a variety of searches for new physics, none of which has 
resulted in any unexpected production.  
The Standard Model is still viable today.  
With many more exciting analyses underway at D\O\ we continue the search.
We thank the organizers of this conference for the opportunity 
to present some of our latest results.

\end{document}